\documentclass[9pt,twocolumn,twoside]{osajnl}
\usepackage[utf8]{inputenc}
\usepackage{amsmath}
\usepackage{physics}
\usepackage{hyperref}
\usepackage{graphics}
\usepackage{graphicx}
\usepackage{amsmath}
\usepackage{lipsum}
\usepackage{mathtools}
\usepackage{cuted}
\journal{ol} 

\setboolean{shortarticle}{true}

\title{Experimental observation of polarization coherence theorem}

\author[1*]{Bhaskar Kanseri}
\author[1]{Sethuraj K. R.}

\affil[1] {Experimental Quantum Interferometry and Polarization (EQUIP), Department of Physics, Indian Institute of Technology Delhi, Hauz Khas, New Delhi-110016, India}

\affil[*]{Corresponding author: bkanseri@physics.iitd.ac.in}

\ociscodes{(260.5430) polarization; (030.0030) Coherence and statistical optics.}

\begin{abstract}
For light fields, the manifestation of correlations between fluctuating electric field components at different space-time points is referred to as coherence, whereas these correlations appearing between orthogonal electric field components at single space-time point are referred to as polarization. In this context, a natural question is: how coherence and polarization are interconnected? Very recently, a tight equality $P^2=V^2+D^2$ namely the ``polarization coherence theorem" (PCT) connecting polarization $P$ with interference visibility $V$ (measure of coherence) and distinguishability $D$ (measure of which-path information) has been proposed [Optica 4, 1113 (2017)]. We here report a direct observation of PCT for classical light fields using a Mach-Zehnder interferometer along with a synthesized source producing a complete gamut of degrees of polarizations. Our experimental demonstration could motivate ongoing experimental efforts towards probing the hidden coherences and complementarity features.
\end{abstract}

\setboolean{displaycopyright}{true}
\setboolean{shortarticle}{true}

\begin{document}

\maketitle

Over the last couple of centuries, optical field correlations played a significant role in understanding several optical phenomena and subsequently leading to innumerable applications in many areas of physics and engineering \cite{wolf03, mandel91, wolf07, brosseau98}. The importance of optical coherence can be envisioned by the fact that even in recent years, several new types of hidden coherences have been discovered and experimentally demonstrated \cite{zela18, svozilik15, eberly16, eberly17}. Complementarity has been studied in several systems in the past decades showing the dual behaviour of quantum particles \cite{zurek79, walther91}. In this context, a natural question is: how quantities related to wave and particle pictures are interconnected? The relationship between interference visibility $V$ (wave picture) and distinguishability $D$ (particle picture) $V^2+D^2 \leq 1$ describes the wave-particle duality of light and has been investigated for many quantum systems either using a Mach Zehnder interferometer or using other interferometers \cite{englert96, rempe98, zela14}. Of late, a tight equality $P^2=V^2+D^2$ connecting polarization $P$ with $V$ and $D$ has been proposed theoretically, termed as ``polarization coherence theorem (PCT)" \cite{vamivakas17}, which quantifies Bohr's complementarity principle \cite{bohr28}. The generalization of this inequality to tight equality $P^2=V^2+D^2$ states that for two conjugate variables, observation of one precludes the observation of other, subject to the degree of polarization of the field. More recently, the intimate connection between entanglement and wave-particle duality has been demonstrated \cite{qian18, jakob10}. The experimental demonstration establishes that the weirdness of quantum theory, i.e. entanglement controls the degree of duality \cite{qian18}.

In this letter, we report an experimental observation of the equality $P^2=V^2+D^2$, thus verifying PCT for the classical polarization states having a full range of variable degrees of polarizations. We use an experimental scheme involving two balanced Mach-Zehnder interferometers; one for the synthesis of the source having a tunable degree of polarization, and the other for the verification of the PCT. In a Mach-Zehnder interferometer, owing to a wide separation of interfering beams, one can control and probe complementary features such as visibility and distinguishability more conveniently compared to the Young's interferometer. We also demonstrate that the equality can be easily derived using Stokes description and coherence matrix treatment of the light fields. We also note that several quantum optical phenomena have been recently revisited from a classical optics viewpoint and it was observed that they exhibit interesting features such as correlations, non-separability, entanglement, Bell violations, etc., with many of them contributing to bring the boundaries between quantum and classical domains closer to each other \cite{eberly11, saleh13, eberly_16, cabello16}. Analogies between coherence, entanglement, and polarization are some of such examples \cite{spreeuw01, spreeuw98, james11, forbes15, leuchs15}. 

\begin{figure}[t]
\centering
\includegraphics[width=\linewidth]{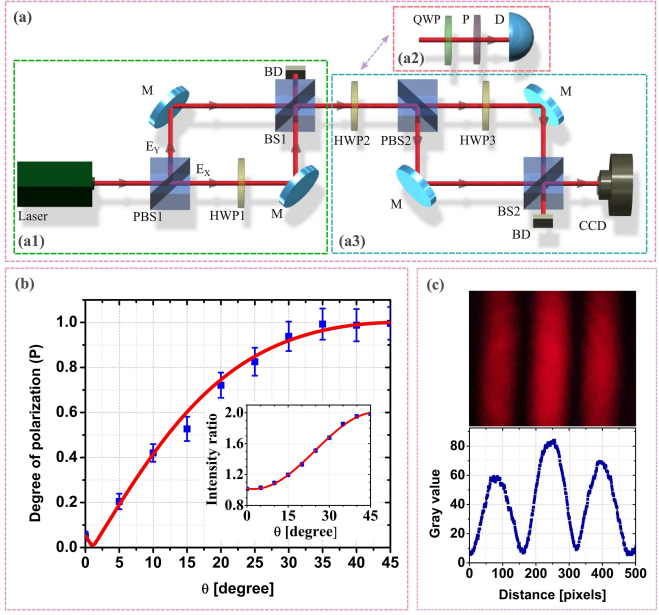}
\caption{(a): Experimental scheme for the observation of polarization coherence theorem. The first interferometer (a1) is used to control the degree of polarization of the field measured using a Stokes measurement system (a2), whereas, the second interferometer (a3) is used to control visibility and distinguishability of interference fringes. Notations: PBS polarization beam splitter, M mirror, HWP half-wave plate, BS beam splitter, BD beam dump, QWP quarter wave plate, P polarizer and D photodetector. (b): Plot showing the degree of polarization ($P$) of the synthesized source as a function of the rotation of half-wave plate (HWP1) angle $\theta$. The blue squares represent the experimental data whereas the solid red line represents the theoretical fit using eq. (3). The inset shows the ratio of reflected and transmitted intensities at the output of the beam-splitter BS1 with respect to the angle of HWP1. The blue squares represents the experimental data, whereas the red line shows the theoretical fit using eq. (2). (c): Upper part shows the interference fringes obtained on the CCD, and the lower part demonstrates the variation of intensity of the interference fringes (upper part) along the horizontal direction, which is used to determine the visibility of fringes [eq. (7)].}
\label{fig:false-color}
\end{figure}

In order to observe PCT in a direct manner, we first construct a source of tunable degree of polarization \cite{kandpal10}. A randomly polarized laser He-Ne laser beam (make Newport, model R30988, wavelength 632.8nm, power 2mW) passes through the first Mach-Zehnder interferometer consisting of a polarizing beam splitter, a half-wave plate (HWP1), and a non-polarizing beam splitter [Fig.1(a)]. For our He-Ne laser, the longitudinal mode spacing is 566MHz which is smaller than the Doppler broadened linewidth of the laser, resulting to several modes lasting simultaneously in the broadened linewidth. Since the adjacent modes are orthogonally polarized, the resultant output beam of the laser is unpolarized. By setting the angle of HWP1, the degree of polarization of the output beam is tuned. The output beam having a known degree of polarization further passes through another half-wave plate (HWP2) which controls the orientation of orthogonal polarization components of the beam entering the second Mach-Zehnder interferometer. Thus for a fully polarized beam, HWP2 can be used to swap the intensities in the second interferometer arms (between the two spatial-modes of polarization). A third half-wave plate (HWP3) with its fast axis at an angle 45 degree with input beam polarization is used to maintain the same polarization state of the interfering beams at the last beam splitter to ensure the maximum visibility of the interference fringes. The interference fringe pattern (image) and intensities of each interfering beams are recorded using a CCD camera [make Olympus, model DP27] to measure the visibility and distinguishability, respectively [Fig.1(c)]. In order to avoid saturation of the CCD, the laser beam intensity was reduced using appropriate neutral density filter after the laser and the integration time of the CCD was accordingly adjusted. For each measurement, 100 data set (fringe) images were taken for statistical analysis and error estimation. In order to reduce the background noise, several background images were taken before capturing the data set images and the mean background image was subtracted from each of the measured data set images. The degree of polarization of the optical field at the output of the first Mach-Zehnder interferometer is determined using Stokes polarimetry (scheme consisting of a set of quarter-wave plate and polarizer) and the light intensity is measured using a homemade photodetector.

Let us consider a light field propagating in $z-$ direction described as $E(t)=\hat{x}E_x(t)+\hat{y}E_y(t)$ passes through the optical path shown in Fig.1(a1), where $E_x(t)$ and $E_y(t)$ are the orthogonal electric field amplitudes. The resultant field at the output of BS1 is given by $\hat{x}[r_xE_x(t)cos2\theta]+\hat{y}[r_yE_x(t)sin2\theta+t_yE_y(t)]$, where $r_x, t_x$ and $r_y, t_y$ are the amplitude reflection and transmission coefficients of BS1 for the horizontal and vertical polarizations, respectively and $\theta$ is the angle of HWP1. Since the orthogonal input field components are uncorrelated, i.e.,$\langle E_x(t)^*E_y(t)\rangle=0$, the polarization (coherence) matrix for the output field after BS1 can be given by (Section 8.1, Eq. 2 of \cite{wolf07})
\begin{equation}
J=
\begin{bmatrix}
r_x^2\langle |E_x(t)|^2 \rangle cos^22\theta & 0.5r_xr_y\langle |E_x(t)|^2 \rangle sin4\theta \\ 
0.5r_xr_y\langle |E_x(t)|^2 \rangle sin4\theta & r_y^2\langle |E_x(t)|^2 \rangle sin^2 2\theta+t_y^2\langle |E_y(t)|^2 \rangle
\end{bmatrix}.
\end{equation}
One can estimate the polarization dependant behavior of a beam-splitter as follows: Using the Jones matrix approach, the ratio of the reflected intensity $I_1(\theta)$ to the transmitted intensity $I_2$ at the output port of the BS1 can be expressed as
\begin{equation}
\frac{I_1(\theta)}{I_2}=C_1cos^22\theta+C_2sin^22\theta,
\end{equation}
where coefficients $C_1=\langle |E_x(t)|^2 \rangle r_x^2/\langle |E_y(t)|^2 \rangle t_y^2 $ and $C_2=\langle |E_x(t)|^2 \rangle r_y^2/\langle |E_y(t)|^2 \rangle t_y^2 $ are proportional to the ratio of reflectivity and transmissivity of BS1 for the horizontal and vertical polarizations, respectively. A plot between the intensity ratio and HWP1 angle for our BS1 is shown in the inset of Fig. 1(b). The experimentally obtained values are fitted with eq. (2), and the ratio parameters $C_1$ and $C_2$ are obtained as 1.01236 and 2.00584, respectively. This confirms the strong polarization dependent behavior of the non-polarizing beam splitter BS1. 

The degree of polarization of the output field for a realistic experimental scheme taking into account the polarization dependence of the beam-splitter is thus obtained as \cite{wolf03}
\begin{equation}
P(\theta)=\sqrt{1-\frac{4det J}{(Tr J)^2}} \equiv \sqrt{1-\frac{4C_1cos^22\theta}{[1+C_1cos^22\theta+C_2sin^22\theta]^2}}.
\end{equation}

The Stokes parameters of the light field can be measured experimentally using the following relations \cite{kandpal10, hauge76}:

\begin{align}
\begin{split}
S_0=I(0,0)+I(0,90),\: S_1=I(0,0)-I(0,90),\\
S_2=I(45,45)-I(45,135),\: S_3=I(0,45)-I(0,135),
\end{split}
\end{align}

where $I(\alpha, \beta)$ represents the intensity detected by the photodetector for the quarter-wave plate and the polarizer making angles $\alpha$ and $\beta$ (in degree) with x-axis, respectively. Then the corresponding degree of polarization (P) can be obtained as ,
\begin{equation}
P=\frac{\sqrt{S_1^2+S_2^2+S_3^2}}{S_0}.
\end{equation}
The experimentally measured value of the degree of polarization obtained using Eqs. (4) and (5) for our practical source using Stokes scheme of Fig. 1(a2) can be varied from 0.05 to 0.99 by changing $\theta$ from 0 to 45 degree, respectively as shown in Fig. 1(b). It was experimentally confirmed by measuring the Stokes parameters \cite{wolf07, hauge76} using a set of quarter-wave plate and polarizer as shown in Fig. 1(a2). The experimental data of Fig. 1(b) fits well with eq. (3) within the experimental uncertainty.

We emphasize here that the second Mach-Zehnder interferometer in our experiment maps the ordinary (spin) polarization to which-path degree of freedom. Thanks to the HWP2 -PBS combination, which splits the orthogonal spin polarizations and guides them to different arms of the interferometer, converting spin- modes to spatial modes of polarization (horizontal ($h$) and vertical ($v$) polarizations in different arms). Polarization generally being a two-party property, these two orthogonal modes can be referred as polarization-spatial modes \cite{vamivakas17, zela18}. Since our experiment deals with partially polarized fields also, the Mach-Zehnder interferometer allows mapping the definition of "partial coherence" to "partial mode coherence". The intensity of each of the polarization spatial-mode is recorded using the CCD, by blocking the other beam. The distinguishability ($D$) can then be determined by measuring contrast between the polarization spatial-modes as

\begin{equation}
D=\frac{|I_h-I_v|}{I_h+I_v}.
\end{equation}

The visibility ($V$) of the interference fringes is obtained by measuring the contrast between the maximum and minimum intensities as

\begin{equation}
V=\frac{I_{max}-I_{min}}{I_{max}+I_{min}}.
\end{equation} 

The fields at the output of the second Mach-Zehnder interferometer (after BS2) can be theoretically calculated using Jones calculus.  Using the expressions for $D$ (Eq. (5)), and $V$ (Eq. (3) of Ref. \cite{vamivakas17}), the final expressions for visibility and distinguishability  after BS2 can be expressed as a function of the angle of HWP1 $(\theta)$ and HWP2 $(\phi)$ (Fig.1) as 

\begin{equation}
D(\theta, \phi)=\left |\frac{cos4\phi[C_1cos^2 2\theta-C_2sin^2 2\theta-1]+\sqrt{C_1C_2}sin4\theta sin4\phi}{C_1cos^2 2\theta+C_2sin^2 2\theta+1}\right |,
\end{equation}
\begin{equation}
V(\theta, \phi)=\left |\frac{sin4\phi[C_1cos^2 2\theta-C_2sin^2 2\theta-1]-\sqrt{C_1C_2}sin4\theta cos4\phi}{C_1cos^2 2\theta+C_2sin^2 2\theta+1}\right |.
\end{equation}

We would like to emphasis that one can explore the ray-wave duality, its relation with the polarization of optical field and with the absolute value of the degree of coherence by making an analysis using the Stokes parameters. Let us consider the light field described in the last section. The Stokes parameters associated with the light field can be written in terms of the elements of polarization (coherence) matrix as\cite{wolf07}
\begin{align}
\begin{split}
S_0=J_{xx}+J_{yy},\: S_1=J_{xx}-J_{yy},\\ 
S_2=J_{xy}+J_{yx},\: S_3=i(J_{yx}-J_{xy}),
\end{split}
\end{align}
where $J_{ij}=\langle E_i^*E_j \rangle$ are the coherence matrix elements. 

Using eq. (10), one can express distinguishability for different polarization bases in terms of Stokes parameters as, $D_{xy}=|S_1/S_0|$, $D_{\pm 45}=|S_2/S_0|$, and $D_{\pm cp}=|S_3/S_0|$; where quantities $D_{\pm 45}$ and $D_{\pm CP}$ correspond to distinguishabilities in diagonal and circular polarization basis, respectively. The resultant intensity $I$ of the interference pattern is given by $I=J_{xx}+J_{yy}+2|J_{xy}|\text{cos}(\text{arg} J_{xy})$ \cite{wolf03}. Using the definition of fringe visibility given as eq. (7) and Stokes parameters from eq. (10), visibilities for different polarization bases yield as $V_{xy}=\sqrt{S_2^2+S_3^2}/S_0$, $V_{\pm 45}=\sqrt{S_1^2+S_3^2}/S_0$, and $V_{cp}=\sqrt{S_1^2+S_2^2}/S_0$. Using these expressions for D and V in different bases, one can readily demonstrate that for any of the given polarization basis, the square sum of distinguishability and visibility comes out to be square of degree of polarization, i.e.  $D^2+V^2=P^2$, which is referred as the \textit{polarization coherence theorem (PCT)} \cite{vamivakas17, zela18}.

\begin{figure}[h]
\centering
\includegraphics[width=9.2cm]{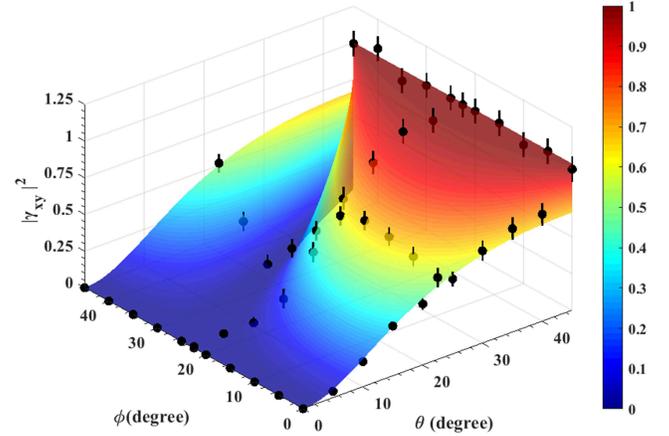}
\caption{Plot of experimental data and corresponding theoretical values [using eq. (12)] for the square of the absolute value of the degree of coherence. The theoretical values are shown for the rotations of half-wave plates HWP1 and HWP2 [see Fig. 1(a)] ranging from 0 to 45 degree. The black spheres show the experimentally measured values of the square of the absolute value of the degree of coherence.}
\label{fig:false-color}
\end{figure}

From the coherence theory of light fields, the absolute value of the degree of coherence is given by
\begin{equation}
|\gamma_{xy}|=\frac{|J_{xy}|}{\sqrt{J_{xx}J_{yy}}}.
\end{equation}

Writing elements of coherence matrix in terms of Stokes parameters using eqs. (8) -(11), one can show that
\begin{equation}
|\gamma_{xy}|^2=\frac{S_2^2+S_3^2}{S_0^2-S_1^2}=\frac{V(\theta, \phi)^2}{1-D(\theta, \phi)^2}.
\end{equation}

\begin{figure}[t]
\centering
\includegraphics[width=1.1\linewidth]{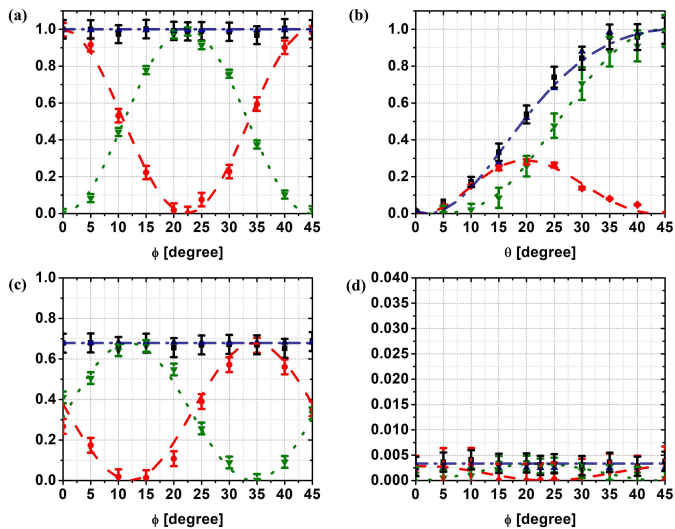}
\caption{Direct observation of polarization coherence theorem (PCT) using experimental scheme of Fig. 1. Plot of the parameters $D^2, V^2$  and $P^2$ for: (a) fully polarized beam for different values of HWP2 angle $\phi$, (b) different degrees of polarization of input field obtained as a function of HWP1 angle $\theta$, (c) partially polarized beam having degree of polarization 0.82 vs different values of HWP2 angle, and (d) for nearly unpolarized beam (within experimental uncertainty), as a function of HWP2 angle. For all figures the represented notations are the following. Red spheres: experimentally measured values of $D^2$, red dashed line:  theoretical fitting of $D^2$ using eq. (8), green inverted triangles: experimentally measured values of $V^2$, green dotted line: theoretical fitting of $V^2$ using eq. (9), blue triangles: $P^2$ values obtained using Stokes measurement, black squares: $P^2$ values calculated as $D^2$+$V^2$, and blue dash-dotted line: theoretical fit of $P^2$ using eq. (3).}
\label{fig:false-color}
\end{figure}

The above expression interprets $|\gamma_{xy}|^2$ in terms of Stokes parameters as well as using the visibility and distinguishability. Clearly, for indistinguishable beams $(D=0)$, $|\gamma_{xy}|$ is equal to the degree of visibility. Using PCT $P(\theta)^2=V(\theta, \phi)^2+D(\theta, \phi)^2$ one can express $|\gamma_{xy}|^2$ in terms of any of the two parameters out of D, V and P. For a field that is unpolarized in the $h-v$ basis $(S_1=0)$, $|\gamma_{xy}|^2$ increases with increase in the degree of polarization in the other two orthogonal bases ($S_2$ and $S_3$). For a partially polarized optical beam, $|\gamma_{xy}|^2$ varies between zero and $P^2$, approaches to zero when $D^2$ approaches to $P^2$, and approaches to $P^2$ when $V^2$  approaches to $P^2$. On the other hand, for the fully polarized case $(P=1)$, irrespective of the value of the degree of distinguishability and degree of visibility, the absolute value of the degree of coherence is always unity. Fig. 2 shows the 3D plot of eq. (12) for the rotation of half-wave plates ($\theta$ and $\phi$). The experimental measurements demonstrate an excellent fit with the theory within the uncertainty of the experiment. This plot confirms the well calibration of our scheme and shows that the polarization of the beam can be manipulated in the desired way.

Using our experimental scheme of Fig.1(a), the values of $V$ and $D$ for the interfering beams of known degrees of polarization are experimentally determined and are shown in Fig. 3. For a nearly fully polarized light ($P\approx1$), orthogonal field components are fully correlated (nearly). One can see that $D$ and $V$ have an inverse relation, i.e., for $D^2\approx1$, any one of the orthogonal field intensities is zero, which signifies $V^2\approx0$. Similarly, for $D^2\approx0$, both of the orthogonal field components have the same intensity, which signifies $V^2\approx1$. For intermediate values of $D$ and $V$ also, the strict condition of PCT is validated. For partially polarized light ($0<P<1$), orthogonal field components are partially correlated. Thus variation in $P$ leads a change in both $D$ and $V$ in accordance with the polarization coherence theorem. The experimental findings confirm this theorem for partially polarized fields also. For a nearly unpolarized light ($P\approx 0$), orthogonal field components are nearly uncorrelated. This infers both $D\approx V \approx 0$, validating PCT for unpolarized field within experimental uncertainty. The experimental observations are confirmed by fitting the experimental data using eqs. (8) and (9), as shown in Fig. 3 and can be seen in an excellent match. 

In conclusion, using a simple interferometric scheme, we have experimentally demonstrated the polarization coherence theorem for classical polarization states of light with a wide gamut of degrees of polarizations. The experimental results fit well with the theoretically expected values. We believe that this direct experimental observation would stimulate experimental research towards revisiting the hidden coherences and various complementarity features, and would also open new doors for understanding the intriguing features such as complementarity and uncertainty in both classical and quantum viewpoints.

\section*{Funding Information}
Science and engineering board (SERB), India (YSS/2015/00743); Council of scientific and industrial research (CSIR), India (03(1401)/17/EMR-II); Board of research in nuclear sciences (BRNS), India (37(3)/20/36/2016-BRNS/37256)


\begin{thebibliography}{1}

\bibitem{wolf03} M. Born, and E. Wolf, \textit{Principles of Optics}, (Cambridge University, 2003).
\bibitem{mandel91} L. Mandel, ``Coherence and indistinguishability," Opt. Lett. \textbf{16}, 1882-1883 (1991).
\bibitem{wolf07} E. Wolf, \textit{Introduction to Theory of Coherence and Polarization of Light}, (Cambridge University, 2007). 
 \bibitem{brosseau98} C. Brosseau, \textit{Fundamentals of Polarized Light: A Statistical Optics Approach}, (Wiley, 1998).
 \bibitem{zela18} F. D. Zela, ``Hidden coherences and two state systems," Optica \textbf{5}, 243-250 (2018).
 \bibitem{svozilik15} J. Svozil\'ik, A. Vall\'es, J. Pe\^rina, Jr., and J. P. Torres, ``Revealing hidden coherence in partially coherent light," Phys. Rev. Lett. \textbf{115}, 220501 (2015).
 \bibitem{eberly16} X.-F. Qian, T. Malhotra, A. N. Vamivakas and J. H. Eberly, ``Coherence constraints and the last hidden optical coherence," Phys. Rev. Lett. \textbf{117}, 153901 (2016). 
\bibitem{eberly17} X.-F. Qian, A. N. Vamivakas, and J. H. Eberly, ``Emerging connections: classical and quantum optics," Opt. Photon. News \textbf{28}, 34-41 (2017).
 \bibitem{zurek79} W. K. Wootters and W. H. Zurek, ``Complementarity in the double-slit experiment: Quantum non-separability and an alternative statement of Bohr's principle," Phys. Rev. D \textbf{19}, 473-484 (1979).
 \bibitem{walther91} M. O. Scully, B.-G. Englert and H. Walther, ``Quantum optical tests of complementarity," Nature \textbf{351}, 111-116 (1991).
 \bibitem{englert96} B. G. Englert, ``Fringe visibility and which-way information: an inequality," Phys. Rev. Lett. \textbf{77}, 2154-2157 (1996).
\bibitem{rempe98} S. D\"urr, T. Nonn, and G. Rempe, ``Fringe visibility and which-way information in an atom interferometer," Phys. Rev. Lett. \textbf{81}, 5705-5709 (1998).
 \bibitem{zela14} F. D. Zela, ``Relationship between the degree of polarization, indistinguishability, and entanglement," Phys. Rev. A. \textbf{89},013845 (2014).
\bibitem{vamivakas17} J. H. Eberly, X. -F. Qian, and A. N. Vamivakas, ``Polarization coherence theorem," Optica \textbf{4}, 1113-1114 (2017).
\bibitem{bohr28} N. T. Bohr, ``The quantum postulate and the recent development of atomic theory," Nature \textbf{121}, 580-590 (1928).
\bibitem{qian18} X. -F. Qian, A. N. Vamivakas, and J. H. Eberly, ``Entanglement limits duality and vice versa," Optica \textbf{5}, 942-947 (2018).
\bibitem{jakob10} M. Jakob, J. A. Bergou, ``Quantitative complementarity relations in bipartite systems: Entanglement as a physical reality," Opt. Commun. \textbf{283}, 827-830 (2010).
 \bibitem{eberly11} X.-F. Qian, and J. H. Eberly, ``Entanglement and classical polarization states," Opt. Lett. \textbf{36}, 4110-4112 (2011). 
 \bibitem{saleh13}  K. H. Kagalwala, G. Di Giuseppe, A. F. Abouraddy, and E. A. Saleh, ``Bell's measure in classical optical coherence," Nat. Photonics \textbf{7}, 72-78 (2013).    
 \bibitem{eberly_16} C. J. H. Eberly, X.-F. Qian, A. A. Qasimi, H. Ali, M. A. Alonso, R. Guti\'errez-Cuevas, B. J. Little, J. C. Howell, T. Malhotra, and A. N. Vamivakas, ``Quantum and classical optics: emerging links," Phys. Scripta \textbf{91}, 063003 (2016).
\bibitem{cabello16} D. Frustaglia, J. P. Baltan\'as, M. C. Vel\'azquez-Ahumada, A. Fern\'andez-Prieto, A. Lujambio, V. Losada, M. J. Freire, and A. Cabello, ``Classical physics and the bounds of quantum correlations," Phys. Rev. Lett. \textbf{116}, 250404 (2016). 
   \bibitem{spreeuw98} R. J. C. Spreeuw, ``A classical analogy of entanglement," Found. Phys. \textbf{28}, 361-374 (1998). 
   \bibitem{spreeuw01} R. J. C. Spreeuw, ``Classical wave-optics analogy of quantum information processing," Phys. Rev. A \textbf{63}, 062302 (2001).
    \bibitem{james11} O. Gamel and D. F. V. James, ``Causality and the complete positivity of classical polarization maps," Opt. Lett. \textbf{36}, 2821-2823 (2011). 
    \bibitem{forbes15} M. McLaren, T. Konrad, and A. Forbes, ``Measuring the nonseparability of vector vortex beams," Phys. Rev. A \textbf{92}, 023833 (2015). 
    \bibitem{leuchs15} A. Aiello, F. T\"oppel, C. Marquardt, E. Giacobino, and G. Leuchs, ``Quantum-like nonseparable structures in optical beams," New J. Phys. \textbf{17}, 043024 (2015).
\bibitem{kandpal10} B. Kanseri and H. C. Kandpal, ``Experimental determination of two-point Stokes parameters for a partially coherent broadband beam," Opt. Comm. \textbf{283}, 4558-4562 (2010).
\bibitem{hauge76} P. S. Hauge, ``Survey of methods for the complete determination of a state of polarization," Proc. SPIE \textbf{88}, 3-10 (1976).
\end {thebibliography}

\newpage
\end{document}